\documentclass{iaus}
\usepackage{graphicx}

\def\solm{M$_{\odot}\,$}

\def\kms{km s$^{-1}$}

\def\solm{M$_{\odot}\,$}

\def\kms{km s$^{-1}$}

\def\mass{$10^{11}$ M$_{\odot}\,$}
\def\hmass{$10^{11.5}$ M$_{\odot}\,$}
\def\lmass{$10^{10}$ M$_{\odot}\,$}

\def\casgm20{CAS-G-M$_{20}\,$}
\def\m20{M$_{20}\,$}

\title[Galaxy Bulges at Mid and High-Redshift] 
{Galaxy Bulges at Mid and High-Redshift}

\author[Christopher J. Conselice]   
{Christopher J. Conselice$^1$}

\affiliation{$^1$School of Physics and Astronomy, University of Nottingham, England\\[\affilskip]}
\pubyear{2007}
\volume{245}  
\pagerange{1-10}
\date{?? and in revised form ??}
\jname{Formation and Evolution of Galaxy Bulges}
\editors{ed. M. Bureau et al.}
\begin{document}
\maketitle
\begin{abstract}

Bulges are a major galaxy component in the nearby universe, and
are one of the primary features that differentiates and defines 
galaxies.  The origin of bulges can be directly probed in part by
examining distant galaxies to search for high redshift bulges, and
to study the properties of bulges in formation.  We review the 
evidence for bulges at high redshift in this 
article, and how by studying bulges through a variety of approaches, 
including morphological, colour, and stellar mass selection,
we can determine when and how these systems assembled.  
We argue that the majority of the most massive `classical' bulges are
in place by $z \sim 1.5 - 2$, and likely formed very early 
through major mergers. Other, likely lower mass, bulges 
form through a secular process along with their disks.  Direct 
observations suggest that these two formation processes are 
occurring, as spheroids are commonly seen at $z > 1$, as are disks and spiral 
galaxies in the form of luminous diffuse objects, clump-clusters, 
and chain galaxies. However, bulge+disk systems are relatively 
rare until $z \sim 1$, suggesting that this structural assembly 
occurred relatively late.

\end{abstract}
\firstsection 
\section{Introduction}

Galaxy bulges are one of the major components of nearby galaxies,
and help define the Hubble classification sequence. Bulges 
are very common, and are found in a large fraction ($>$ 50\%) of all 
nearby galaxies.   Depending on how bulges are defined, a large 
fraction of the light, and stellar mass, in nearby galaxies are 
contained within these systems (e.g., Benson et al. 2002).  
Understanding the formation modes and history of bulges is therefore
clearly important for our understanding of all galaxies, as 
well as for uncovering how and when the Hubble sequence assembled.

This paper reviews our current understanding of mid- and 
high-redshift galaxy bulges and/or the progenitors of modern galaxy bulges.  
Since galaxy bulges are fairly common in the local universe, we should  
be able to explore in a fairly straightforward manner the formation 
and evolution of these systems by looking at more distant galaxies.  We 
describe in this review the current
evidence for the existence of bulges at higher redshifts, based on 
observations of spheroids and spirals/disks, and how the physical 
evolution of these galaxies suggest how bulges are assembling.

Before we can study bulges in the early universe we must be clear
about the definition of what makes a bulge, and how to locate their
progenitors.  This is a primary question to address, but one in which 
there is no generally agreed upon answer. In this review we examine the 
evolution of bulges using several techniques and methods. One method is
to assume that a bulge must be attached to a disk galaxy, and 
that all large, massive disk galaxies have a bulge, or a proto-bulge, at their
centre. The other working definition is to assume that a bulge is a spheroid 
that acquired a disk through an accretion process, or disk assembly.   
Studying bulge progenitors thus requires that we identify massive or 
spheroid galaxies at early times. We can then determine when 
the most massive bulges formed their stellar mass, if not already their 
structure/morphological state.  We argue that many very massive 
bulges are formed by $z \sim 1.5 - 2$, based on the number
densities of massive galaxies at early times.

The other approach for finding bulges at high redshift is to search for 
disk galaxies in formation. This can be effectively done by searching 
for what appears through quantitative and qualitative methods
to be disk galaxies at $z > 1$.  Features used to find these disk
galaxies include: spiral arms, bars, star forming knots, and exponential
or shallower light profiles. These proto-disks include the morphologically
and luminosity defined: luminous diffuse objects (LDOs), clump-clusters, 
and chain galaxies.  While normal appearing disk plus bulge systems are rare 
at $z > 1$, there are many disk-like candidates, and bulge candidates, at 
higher redshifts whose properties can be studied. Observations 
show that these galaxies are undergoing a significant amount of star 
formation,
and are often composed of multiple clumps of young stars that are
potentially producing a bulge through secular evolution.

The current evidence at high-redshift suggests that more than one 
mechanism is responsible for the creation of bulge+disk systems. While
spheroids and disk-like galaxies can be found at $z > 1$,
systems which could be identified as bulge+disks
are relatively rare at these epochs, suggesting that the full
Hubble sequence, and disk+bulge galaxy morphologies, are established 
relatively late.  Throughout this paper we use the standard cosmology
of H$_{0} = 70$ km s$^{-1}$ Mpc$^{-1}$, and 
$\Omega_{\rm m} = 1 - \Omega_{\Lambda}$ = 0.3, and a Chabrier
IMF for stellar mass measurements, unless otherwise noted.

\section{Identifying High-Redshift Bulges}\label{sec:spheroids}

\subsection{Nearby Bulges: A $z = 0$ Benchmark}

Galaxies bulges are well studied in the nearby
universe (e.g., Kormendy \& Kennicutt 2004), and observations of nearby
disk+bulges continue to reveal important information on their
formation (e.g., Lanyon-Foster et al. 2007; Drory \& Fisher 2007).
Before we can effectively explore the formation and presence of
mid- and high-redshift bulges it is important to quantify the
number of these systems in the nearby universe to serve
as a bench-mark for the maximum number of systems we expect to find at
earlier times. This can be done in a number of ways, including 
investigating the number of bulges, or galaxies which host bulges, in 
the nearby universe.  

Within the nearby universe, bulge-dominated galaxies are the most
common galaxy type.  Galaxies classified as early-type disks are found with 
a number density of $\sim 1.3\pm0.3 \times
10^{6}$ Gpc$^{-3}$ h$^{3}_{100}$, with a similar number of late-type 
disks (Conselice 2006a), which also generally host bulges.   Based 
on these densities we should easily find and detect bulges, or
bulge progenitors, at higher redshifts.  For example, between 
$0.2 < z < 1$ there should be 20 galaxies with massive bulges in the 
Hubble Deep Field, 1000 within the GOODS-fields, 2000 within the 
Extended Groth Strip, and 20,000 within the COSMOS field.  
A similar number of bulges/progenitors should be 
found at higher redshifts in these fields.  Therefore, there is no lack of 
opportunity, even within relatively small Hubble Space Telescope imaging
areas, to locate and study bulges at higher redshifts.

\begin{figure}
\hspace{-0.5cm}
\includegraphics[height=2.75in,width=5.5in,angle=0]{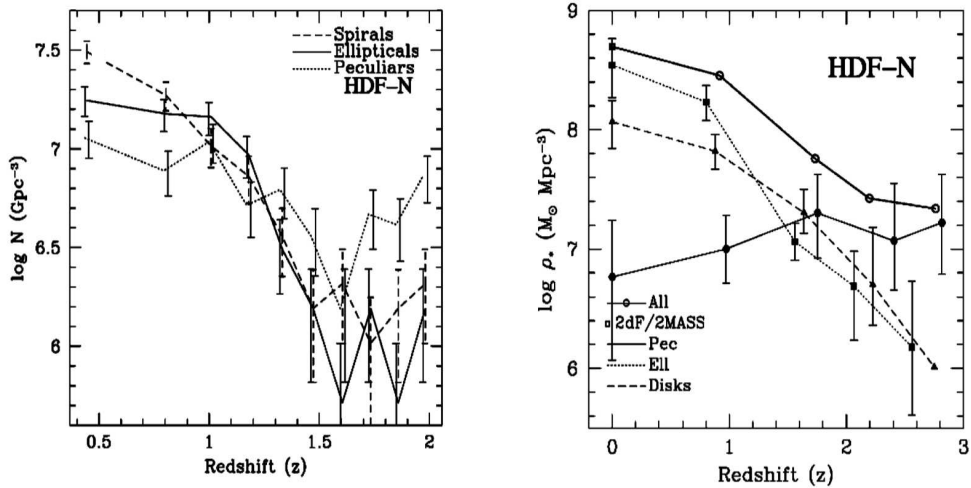}
  \caption{The relative contributions of various galaxy types to
the galaxy population and the integrated stellar mass density
as a function of redshift.  The left panel shows the number
density evolution for galaxies at $I < 27$ up to $z \sim 3$.
The right panel shows the evolution in the fraction of the galaxy
stellar mass density associated with galaxies of various morphological
types. }\label{fig:contour}
\end{figure}

\subsection{High Redshift Bulges as Spheroids}

The dominate paradigm of galaxy formation is based on the
successful $\Lambda$-dominated cold dark matter theory of structure
formation.  In this picture, bulges, and spheroids in general,
are the oldest galaxy components in the universe, and  formed
before the disk galaxies we see today.  Within this formation scenario, 
which still lacks much observational basis, these spheroids and bulges
are formed through the mergers of primordial galaxies which 
are rotationally supported (e.g., Baugh et al. et al. 1998).

If galaxy bulges form in an inside-out fashion, such that the
bulge is present before the disk, then looking for galaxy spheroids
at mid- and high-redshift is an excellent way to study the progenitors
of galaxy bulges, and how they form.  There are several ways to study 
distant spheroids that may be the progenitors of bulges in
modern disk galaxies.  The most obvious examples are to search for 
spheroids morphologically, to search for red bright galaxies, and to locate
massive galaxies at high redshift.

The first way to approach this problem is to determine how galaxy
morphology evolves with cosmic time. Currently, the only field where
this can be carried out, without having to account for morphological
k-corrections, is within the Hubble Deep Field-North (HDF-N) (Conselice
et al. 2005a).    Figure~1 shows the evolution of the relative
contributions of spheroids, disks and peculiars to the galaxy population
as a function of redshift. As can be seen at $z < 1$,
spheroids and disks are the dominate galaxy population, while peculiar
galaxies are the most prominent at higher redshifts.  

Based on this, we can conclude that while the stars that make up bulges
might be present at high redshift ($z > 1$), these systems are not in 
their nearby
relaxed morphological state, at least not in the same number that they
are found at $z < 1$.  This implies that at least some evolution is occurring
within bulges at $z > 1.5$, and that very few, if any, disk+bulge systems
are present at $z > 1.5$. The final structural assembly of disks with
bulges thus occurs relatively late in the history of the universe.   While the 
HDF-N is a small
field of view, other studies have found a similar lack of conclusive
examples of bulges+disk
systems at higher redshifts (e.g., Conselice et al. 2003a, 2005a; Papovich
et al. 2005; Ravindranath et al. 2006; Conselice et al. 2007a).

While bulge+disk systems are not present in a high abundance at
higher redshifts, it is possible and as we shall argue, likely, that
bulge progenitors are present at $z > 1$.  
If we examine galaxy bulges as massive galaxies, then we can measure 
how much of the stellar mass within massive galaxies has formed 
at relatively high
redshifts.  This was recently examined by Conselice et al. (2007b) who
found that the majority of massive galaxies with M$_{*} >$ \mass are
present by $z \sim 1$, while those of very high mass, with M$_{*} >$ \hmass
are statistically present at their $z \sim 0$ number and mass densities 
by $z \sim 2$ (Figure~2) (see also e.g., Fontana et al. 2004; 
Daddi et al. 2004).

We can use Figure~2 to argue that most of the stellar mass in the most
massive bulges was likely in place by $z \sim 1.5-2$. The reason for this 
is that galaxies with masses M$_{*} >$ \mass are nearly all present by
these redshifts, and in the local universe  roughly 40\% of
all galaxies with M$_{*} >$ \mass are disk galaxies (Conselice
2006a).  This implies that a fairly large fraction of the massive
galaxies we see at high redshift are the progenitors of bulge like
galaxies.  While most of these massive galaxies do not have
disks surrounding them, we find that a fraction however are within
spiral like galaxies (Conselice et al. 2007b).

\begin{figure}
\hspace{-0.5cm}
\includegraphics[height=2.55in,width=5.5in,angle=0]{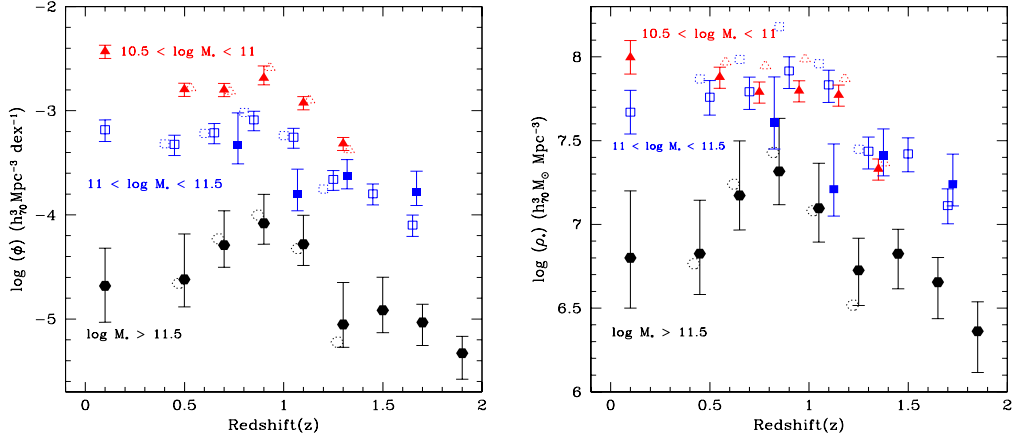}
  \caption{Left panel: the evolution in number densities for galaxies of
various masses between $z \sim 0.4 - 1.4$.  Right panel: the
stellar mass density evolution as a function of galaxy mass at the same
redshift intervals.   The error bars listed on both the
number and mass densities reflect uncertainties from stellar mass
errors, as well as cosmic variance, and counting statistics.}\label{fig:contour}
\end{figure}


\subsection{High Redshift Bulges as Disks}

Another way to search for high-redshift bulges is to find examples of 
bulges+disks, or pure disk galaxies, that might be forming bulges 
at $z > 1$. This is an important exercise to carry out as it allows us to 
directly measure when, and perhaps how, the Hubble sequence was put 
into place.
It also lets us determine whether bulge formation is an
outside-in process, or an inside-out process, since if we found significant
numbers of disks in formation that did not yet have bulges this would imply
that the bulges form after their disks are in place.

The evidence for disk galaxies at high redshift is limited, and in
some cases ambiguous.
There are several reasons for this, including 
that disk galaxies are probably different morphologically in the past
than they are today,  making them difficult to identify. Also, the
majority of high resolution Hubble imaging probes the rest-frame ultra-violet
at $z > 1.5$, making if difficult to identify disks+bulges (e.g.,
Windhorst et al. 2002; Taylor-Mager et al. 2007). There are
however many examples of disk galaxy candidates at high redshifts
that are almost certainly the progenitors of the disk galaxies we
see today.

\begin{figure}
\hspace{-0.5cm}
\includegraphics[height=4in,width=6in,angle=0]{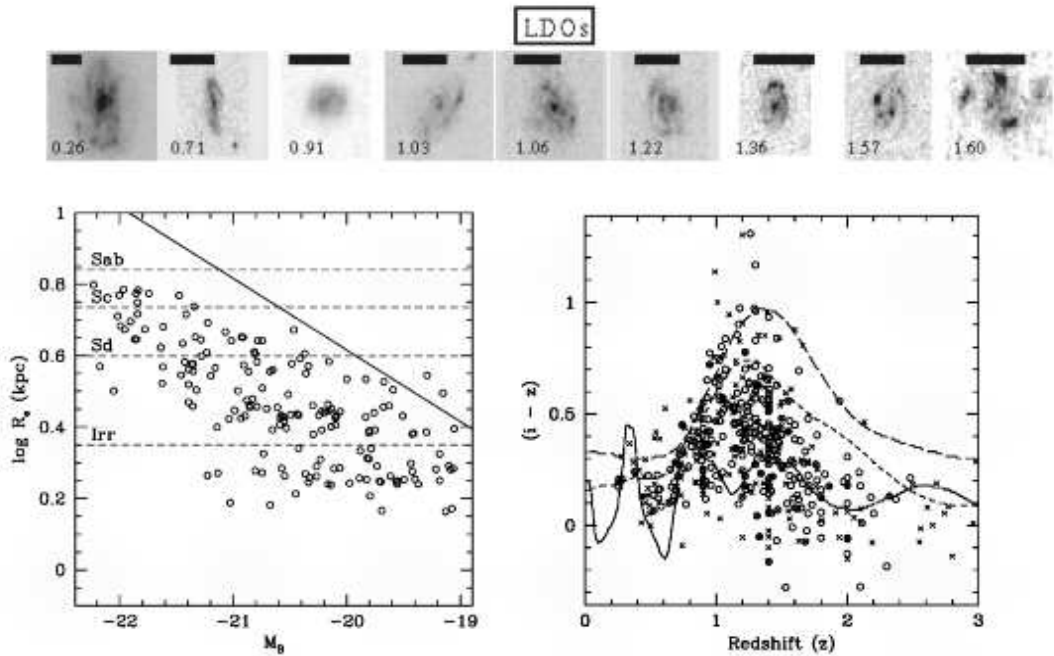}
  \caption{Top: images of luminous diffuse objects (LDOs) at $z < 1.5$.
Left: the size-absolute luminosity relation for the LDOs with the sizes
of nearby galaxy types labelled and the relation for nearby disks shown
as a solid line.  Right: the observed $(i-z)$ colours for LDOs, showing
a diversity of colour at different redshifts.  The lines are, from bluest
to reddest: starburst (solid line), Scd (dashed), and Sbc (long dashed).}\label{fig:contour}
\end{figure}

The first instance of large disk galaxies seen at high redshift
originated from deep Hubble Space Telescope imaging in the 
Hubble Deep Field South (Labbe et al. 2003), and the GOODS 
fields (Conselice et al. 2004).  These early studies found
examples of galaxies that appear to have disk-like features, 
such spiral arms, bulges, bars, and clumpy
star forming regions.  Conselice et al. (2004) took this analysis 
further by investigating
a population of possible proto-disk galaxies called luminous
diffuse objects (LDOs) which are located by their low light concentrations,
high luminosities, and large sizes (Figure~3). These galaxies are 
found in abundance at $z > 1$, with most systems at $1 < z < 2$.  The 
size-luminosity
relation for the LDOs is similar to nearby disk galaxies, but with a
luminosity offset of a few magnitudes. The spectral energy distributions
of these galaxies also suggests that they are undergoing starbursts
(Figure~3).

Proto-disk galaxies have furthermore been studied in great
detail by e.g., Elmegreen et al. (2007, and references therein).  
These studies find that a large fraction of all galaxies
at $z > 1$ are composed of clumpy light made up of very young 
($\sim 100$s Myr) star clusters, which have masses ranging
from $10^{7} - 10^{9}$ \solm. These clusters are found in both
the classical chain galaxies, as well as in the clump-clusters.
As shown by Elmegreen et al. (2004) these two galaxy types
are taken from the same population, with chain galaxies the edge-on 
version of clump-clusters.  The evidence for this includes the fact
that these two galaxy populations
have similar sizes, magnitudes, and clump-cluster properties, as well
as a combined axial ratio distribution expected for
a disk population (Elmegreen \& Elmegreen 2005).  Furthermore,
more recent NICMOS observations of clump-clusters and LDOs
suggest that their massive clusters are intrinsic to the galaxy, and are not
lower mass star forming knots (Elmegreen et al. 2007).

\begin{figure}
\hspace{0cm}
\includegraphics[height=6.5in,width=5.5in,angle=0]{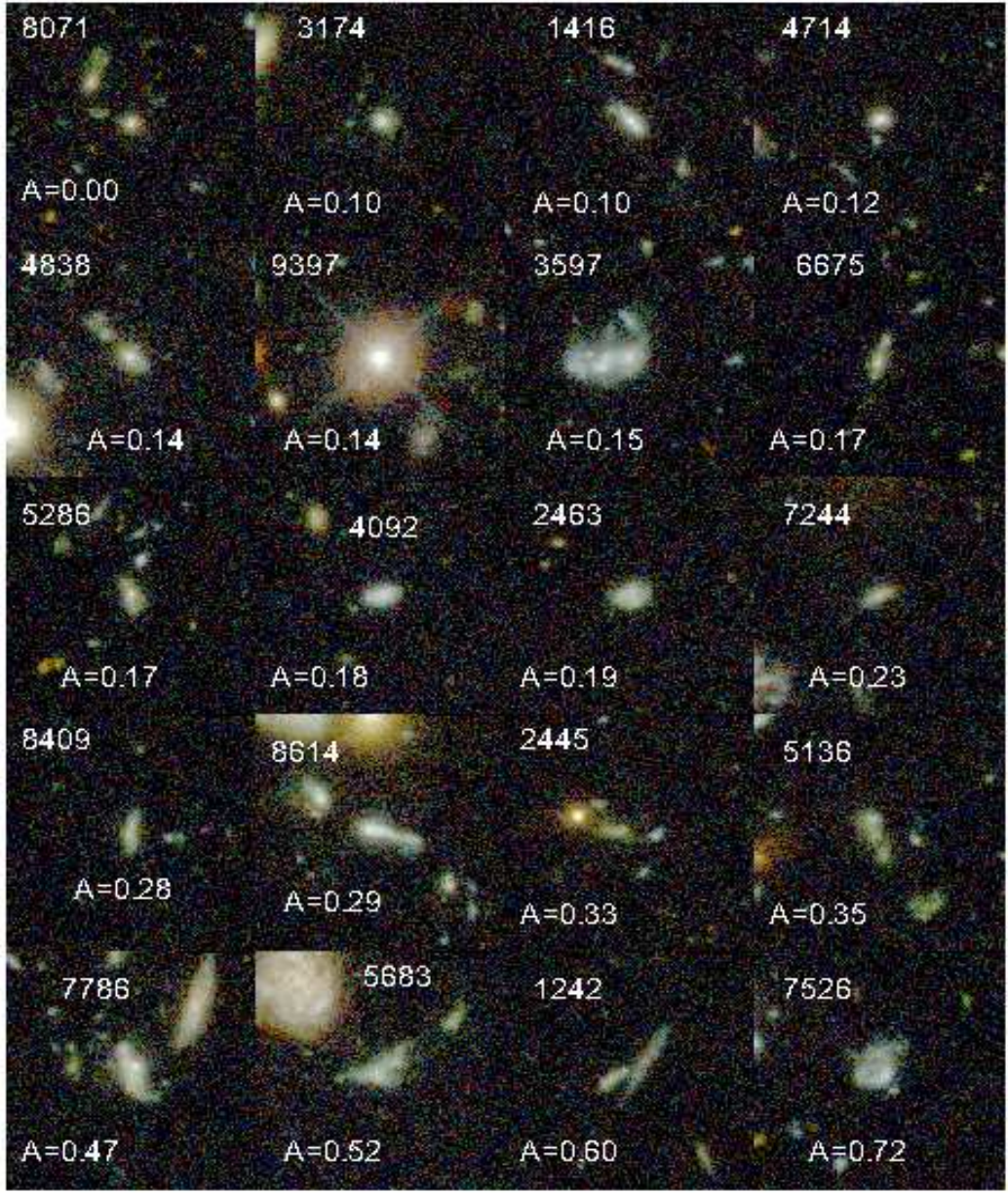}
  \caption{Images of high redshift galaxies as seen in the Hubble Ultra
Deep Field.  This example shows images of galaxies at $2.2 < z < 3.0$ 
and with stellar masses M$_{*} >$ \lmass. The bottom number is the asymmetry
for each system, and the upper number is its ID. }\label{fig:contour}
\end{figure}

\subsection{Kinematics of Disks at $z > 0.5$}

Most of the above discussion relates to the morphological
properties of distant galaxies.  Another area where there
is growing information concerning disk galaxies at mid- and 
high-redshifts is through the use of internal kinematics.  
For example, based on
rotation curves, it is possible to find morphologically selected
disks with well ordered rotation up to $z \sim 1.2$, and
to study the evolution of disk galaxy scaling relations
(e.g., Vogt et al. 1997; Bohm et al. 2004; Ravindranath et al.
2004).

Detailed analyses of mid-redshift disk galaxies suggest that the
formation of disks at $z < 1.2$ is hierarchical.  We know
that disk galaxies at $z < 1.2$ are undergoing 
star formation (e.g., Wolf et al. 2005), and based on their sizes, it appears
that the largest disk galaxies already have their maximum
size, and roughly $z \sim 0$ morphological properties by $z \sim 1$ 
(e.g., Jogee et al. 2004; Ravindranath et al. 2004).  Thus,
it might appear that the star formation within disks
is occurring due to a conversion of existing gas into
stars through the existing disks.  

However, by comparing the stellar and total masses
of disk galaxies at $0.2 < z < 1.2$, Conselice et al. (2005b) 
argue that disk galaxies grow hierarchically during this epoch. 
This can be seen by the fact that the ratio of disk galaxy stellar to 
total mass does not evolve at $z < 1.2$.
If the gas which star formation at $z < 1.2$ arises
is present within disks at $z \sim 1.2$, then the ratio of stellar to
total mass would grow with time as gas is converted into
stars.  Since this ratio of masses is
constant, it implies that there is gas and dark matter accretion 
onto disks during this time, and thus disk formation is
largely hierarchical at $z < 1.2$.

This observation is for disk galaxies that are
mature morphologically at $z < 1.2$,  and does not address 
how disk galaxy formation occurs at earlier times.  While
we have discussed several examples of morphologically selected
disk galaxies at higher redshifts, there are now several
studies which have found evidence for rotating disks in morphologically 
peculiar galaxies at $z > 1.5$ using near-infrared long-slit and integral 
field spectrography.

Based on these observations, there 
are several examples of galaxies which appear to
be rotating disks at these redshifts, with rotation rates
$> 200$ \kms.  Many of these galaxies also show
a large velocity dispersion, and some morphologically
appearing `disks' do not have any obvious rotation along
their major axis (e.g., Erb et al. 2004), although the
limited spatial extent of these observations makes it difficult
to predict how a rotating disk would appear,
especially if still in formation (e.g., Bournaud et al. 2007).

Integral field spectroscopy has however found that the majority of 
star forming selected galaxies at $z \sim 2$ have velocity
gradients of up to 400 \kms (e.g., Forster Schreiber et al. 2004;
Bouche et al. 2007).   Some of these systems
have morphologies and kinematics that suggest they are
disks, while it is difficult to distinguish rotation
from galaxy merging in other systems (Forster Schreiber 
et al. 2004).  There are more certain examples of 
large-scale rotating disks, using adaptive optics observations, 
such as the system at $z = 2.38$ studied by Genzel et al. 
(2006), who find a disk galaxy with V$_{\rm circ} = 230$ \kms, 
size 4.5 kpc, and a total mass of $\sim 10^{11}$ \solm.

\section{Bulge Formation Mechanisms}

\subsection{Galaxy Merging}

While it is generally believed that bulges 
may have different formation mechanisms (e.g., Kormendy
\& Kennicutt 2004), with the classical 
bulges forming through mergers, and the `pseudo-bulges' forming
through secular processes, this has not yet been directly shown
or argued using high redshift data.

What we do know from \S 2.2 is that most of the stellar mass in
the most massive galaxies is largely in place by $z \sim 1-2$, and it
is likely that this epoch marks the end of the classical bulge
formation phase, and we must probe higher redshifts to determine how
massive bulges form. 

One of the ways to do this is to investigate the merger fraction,
and merger rate, for massive galaxies, with M$_{*} >$ \lmass
at $z > 2$  (Conselice et al. 2003a; Conselice et al. 2007c, in prep; Figure~4
shows example Hubble images of these systems).  Through
the quantitative CAS structural parameters (e.g., Conselice et al. 2000a,b;
Bershady et al. 2000; Conselice et al. 2002; Conselice 2003) we can 
identify major mergers at mid- and high-redshifts
(e.g., Conselice et al. 2003a,b; Conselice 2006b; Pope et al. 2005; Bridge 
et al. 2007). What these
studies have found is that the most massive galaxies, for which
we have rest-frame optical morphologies, appear peculiar at
$z > 2$ (Conselice et al. 2005a), and are likely in a major
merger phase (e.g., Conselice et al. 2003a, 2007c).

By examining the evolution of the merger fraction with time, and
by using derived time-scales for merging from N-body simulations
(Conselice 2006b), we can determine the merger
history for massive galaxies at $z < 3$.  The result of this, using a 
combined
Hubble Ultra-Deep Field and Hubble Deep Field-North
sample, is shown in Figure~5. These results show that the most massive 
galaxies with M$_{*} >$ \lmass at $z \sim 2.5$
undergo on average 4-5 major mergers at $z < 3$, but that most of this merging
occurs at $z > 1.5$.  This suggests that the most massive galaxies,
and thus the most massive bulges, formed in major mergers
at $z > 2$.

\begin{figure}
\hspace{-0.5cm}
\includegraphics[height=2.75in,width=5.5in,angle=0]{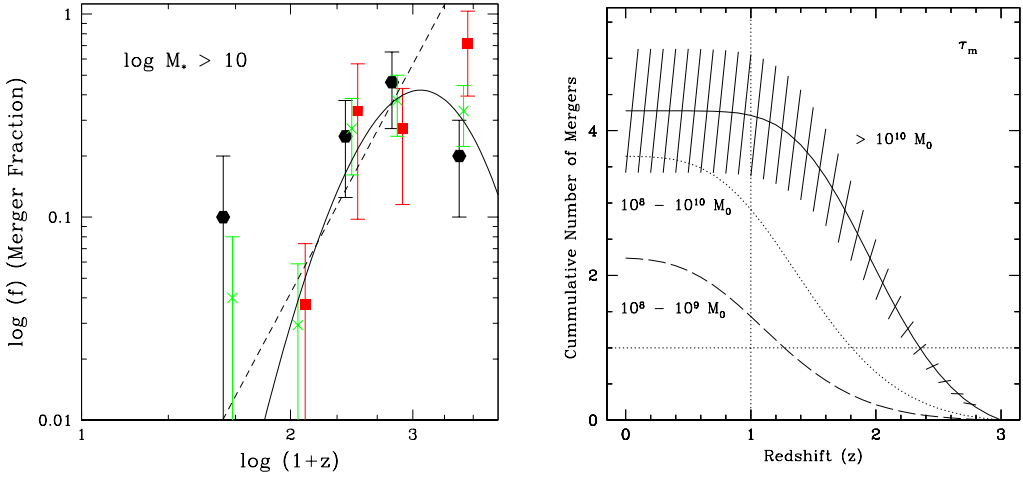}
  \caption{Left panel: The merger fraction for galaxies with
stellar masses M$_{*} >$ \lmass up to $z \sim 3$ in the 
Hubble Ultra-Deep Field (red boxes) and the Hubble Deep Field-North
(solid circles) and a combined sample (green crosses).
Right panel: the evolution of the cumulative number of mergers
occurring within galaxies at $z < 3$ (Conselice et al. 2007, in
prep). }\label{fig:contour}
\end{figure}

\subsection{Disk Formation}

It is generally agreed that disks themselves cannot form through
the merger process, at least in mergers that do not include significant
amount of gas that is ejected and falls back onto a spheroid. In general, 
to form a disk, it is necessary to accrete gas
from the intergalactic medium onto an existing bulge, or onto itself,
to form a pure disk galaxy.  This process appears likely to be the formation 
mechanism for at least some of the clump-clusters or LDOs which
do not have any obvious bulges.

This formation process has been investigated by several teams
who have simulated how a rotating disk made purely of gas will form stars
(e.g.,  Noguchi 1999; Immeli et al. 2004; Bournaud et al. 2007). 
All of these studies have found that a gas rich rotating disk
will fragment into large star clusters throughout,
producing a morphology similar to the LDOs, clump-clusters, and chain galaxies.
The evolution of these disks is driven largely by the interaction of
these star clusters.  In simulations these clusters lose angular
momentum, dissolve, and disturb the underlying disk to such a degree
that they are able to form a bulge in the centre of the disk, and
later an exponential-like profile for the disk itself (Bournaud
et al. 2007).

These simulations are also able to address the life-time of this clump-cluster 
phase within these systems. The latest simulations from
Bournaud et al. (2007) suggest that this cluster phase lasts for
roughly 0.5 Gyr, and only occurs once in the life-time of a disk.
Before this, the galaxy is not forming stars, and after this
phase, the galaxy appears as a disk+bulge system within the simulation.

Using high resolution Hubble imaging, we can determine directly
how the properties of bulges and disks evolve at $z < 1.2$, where
both of these components are found in the same galaxy. An early 
study by Ellis et al. 
(2001) found that the bulges of disk galaxies at $z < 1.2$ are bluer than 
pure spheroids at similar redshifts.  More recently, Koo et al.
(2005) found the opposite conclusion - that the bulges of
spiral galaxies are red, and likely as
old as spheroids at similar redshifts.  Both studies however find that the 
disks of these distant systems are either bluer, or have similar colours, 
suggesting that bulges are older than their disks, or more likely
that disks have had a more recent episode of star formation.  However,
not all `spheroids' are old, as there are examples undergoing star 
formation at $z < 1$ (e.g., Stanford et al. 2004; Teplitz et al. 2006).

\section{Conclusions and Future Outlook}

The study of bulges at mid- and high-redshifts is a very active
field which will advance greatly during the next few years with
high resolution near-infrared imaging with ground-based adaptive
optics and with WFC3 on Hubble.  We can however draw some conclusions 
about the way that bulges form based on current observations of the 
distant universe.

Observations show that there are many disk galaxies, as selected 
morphologically, up to $z \sim 1$, with a similar number density as
disks today. However, at higher redshifts disk galaxies, and in
particular disk galaxies with bulges, are rare.  Most galaxies at
$z > 1.5$ appear peculiar in appearance. Therefore, to study bulges
at $z > 1$ we must search for bulge progenitors. If we  consider bulges 
as  massive spheroids, we find that most massive 
M$_{*} >$ \mass galaxies are formed by $z \sim 1.5 - 2$. Since 
roughly 40\% of M$_{*} >$ \mass galaxies at $z \sim 0$ are massive 
disk galaxies, this suggests that a large fraction of the stars in 
the most massive bulges are in place by this time.

However, finding bonafide disks forming around spheroids is difficult,
and few convincing massive bulge+disks systems have been identified
at $z > 1.5$ (e.g., Figure~4).  This is likely partially
due to the lack of wide-area high-resolution
near-infrared imaging of distant galaxies.  There are however many 
examples of disk galaxies that appear to be forming through a 
secular process, including perhaps the establishment of a bulge.  
Examples of these are the clumpy galaxies at  $z > 2$ identified as
luminous diffuse objects (LDOs), clump-clusters, and chain galaxies,
that are likely a subset of the modern disk galaxy population in
formation.  Simulations show that the clumps in these galaxies interact 
over a few 100 Myr to form a bulge (e.g., Bournaud et al. 2007).

In the future, imaging observations with adaptive optics and the WFC3 on HST
and integral field spectroscopy in the NIR will greatly advance the
study of galaxy bulges at mid- and high-redshifts. What is not clear yet
is the relative role of the clump-cluster vs. gas accretion on massive
spheroids to form bulge+disk systems.   Identifying disks and bulges
in formation through kinematics and high-resolution NIR imaging at $z~>~1$
will ultimately reveal how bulges are established in disks, providing
complementary information obtain from nearby bulges.

\end{document}